\documentclass[a4paper,11pt]{article}
\usepackage[pdftex]{graphicx}
\usepackage[T1]{fontenc}
\usepackage{lmodern}
\usepackage{slashed}
\usepackage[utf8]{inputenc}
\usepackage[english]{babel}
\usepackage{microtype}
\usepackage{cite}
\usepackage{amsmath}
\usepackage{amssymb}
\usepackage{amsfonts}
\usepackage[nottoc,notlot,notlof]{tocbibind}
\usepackage{upgreek}
\usepackage{empheq}
\usepackage{mathtools}
\numberwithin{equation}{section}
\allowdisplaybreaks
\usepackage[all]{xy}
\usepackage{color}
\usepackage{graphicx}
\graphicspath{{images/}}
\usepackage{geometry}
\usepackage[toc,page]{appendix}
\usepackage{hyperref}
\usepackage{tikz-cd}
\usepackage{float}
\usepackage[normalem]{ulem}

\usepackage{bm}
\usepackage{ragged2e}
\usepackage{appendix}
\usepackage{slashed}
\usepackage{bbold}
\usepackage{cancel}
\usepackage{braket}

\usepackage{amsthm}

\newtheorem*{theorem*}{Theorem}

\DeclareMathAlphabet{\mathpzc}{OT1}{pzc}{m}{it}

\definecolor{blue-violet}{rgb}{0.54, 0.17, 0.89}
\definecolor{PineGreen}{cmyk}{0.92, 0, 0.59, 0.25}
\definecolor{YellowOrange}{cmyk}{0, 0.42, 1, 0}
\definecolor{orange}{rgb}{0.95, 0.5, 0.1}

\begin{document}
\begin{titlepage}
\begin{flushright}
LMU-ASC 09/25\\
\par\end{flushright}
\vskip 1.5cm
\begin{center}
\textbf{\Large \bf On Spinning Particles, their Partition Functions \\ \vspace{.2cm} and Picture Changing Operators}
\vskip 1cm
\vskip 0.5cm
\large {
\bf 
E.~Boffo$^{~a,}$\footnote{boffo@karlin.mff.cuni.cz}, P.~A.~Grassi$^{~b,c,}$\footnote{pietro.grassi@uniupo.it}, 
O. Hulik$^{~d,}$\footnote{ondra.hulik@gmail.com}, 
and I. Sachs$^{e,}$\footnote{Ivo.Sachs@physik.uni-muenchen.de}
}  
\\
~
\\
{$^{(a)}$ \it Faculty of Mathematics and Physics, Mathematical Institute, \\ 
 Charles University Prague, Sokolovsk\'{a} 83, 186 75 Prague}
\\
{$^{(b)}$ \it Dipartimento di Scienze e Innovazione Tecnologica (DiSIT),} \\
{\it Universit\`a del Piemonte Orientale, viale T.~Michel, 11, 15121 Alessandria, Italy}
\\
{$^{(c)}$ \it INFN, Sezione di Torino, via P.~Giuria 1, 10125 Torino, Italy}
\\
{$^{(d)}$ \it Institute for Mathematics 
 Ruprecht-Karls-Universitat Heidelberg,
\\ 69120 Heidelberg, Germany}
\\
{$^{(e)}$ \it Arnold-Sommerfeld-Center for Theoretical Physics, Ludwig-Maximilians-Universit\"at \\ Theresienstr. 37, D-80333 Munich, Germany}

\end{center}
\vskip  0.2cm
\begin{abstract}
We compute the partition function for the $N=1$ spinning particle, including pictures and the large Hilbert space, and show that it counts the dimension of the BRST cohomology in two- and four-dimensional target space. We also construct a quadratic action in the target space. Furthermore, we find a consistent interaction as a derived bracket based on the associative product of world line fields, leading to an interacting theory of multiforms in space-time. Finally, we comment on  the equivalence of the multiform theory with a Dirac fermion. We also identify the chiral anomaly of the latter with a Hodge anomaly for the multiform theory, which manifests itself as a deformation of the gauge fixing.


\end{abstract}
\vfill{}
\vspace{1.5cm}
\end{titlepage}
\newpage\setcounter{footnote}{0}

\tableofcontents

\section{Introduction}

Spinning particles are a relatively old subject that still attracts theoretical and mathematical physicists. A non-exhaustive and subjective list of articles focusing on the -quantum- physical states of the theory is \cite{Brink:1976uf,VanHolten:1988jy,Getzler:2015jrr,,Bastianelli2005}, while for their use in path-integral calculations and backgrounds of the target space field theory we refer to  \cite{Bastianelli:2023oca,Bonezzi:2020jjq,Howe89,HOWE1988555,Cremonini:2025eds} for an incomplete list.

The $N=1$ spinning world line for a relativistic particle is generally considered as representing a space-time fermion with the world-line fermions, $\psi^\mu$  acting as Dirac matrices on the representation space generated from some highest weight state for a Clifford algebra. However, it does also have an interpretation as a theory of multiforms in the target space-time \cite{Brink:1976uf,Cremonini:2025eds}, where $\psi^\mu$ represents a basis of 1-forms. However, as a result of the Clifford algebra of $\psi^\mu$, these differentials are endowed with a Clifford multiplication instead of the wedge product. This is why the form degree of the space-time fields is not filtered, giving rise to a theory of multiforms. 

In the absence of extra structure, the world-line describes the propagation of free particles. We will analyze their spectrum in terms of a Hilbert series and compare it with an explicit construction of the cohomology. This is an extension of the discussion in \cite{Boffo:2024lwd}, where the $N=2$ particle was analyzed in this respect. Gauge fixing of the super reparametrization invariance on the world line results in a graded Lie algebra of the corresponding ghost system, which allows for different inequivalent representations (pictures). In \cite{Boffo:2024lwd}, we explored all possible pictures and demonstrated that the techniques initially developed for superstrings (see \cite{Friedan:1985ge}) apply seamlessly in this context so that this approach led to a precise identification of all sectors of the theory. 

In this work, we take a step back and reanalyze the $N=0$ and $N=1$ superparticle using the same methodology as in \cite{Boffo:2024lwd}. We establish the correspondence between the partition function, the light-cone gauge fields, and the covariant expression for the cohomology of the BRST charge. 

For the $N=0$ world line, we then rederive the target space action within the BRST framework, constructing the BV bracket and the Hodge dual operator acting on the Hilbert space of states. While the result is well-known, we generalize it by introducing an arbitrary potential together with the kinetic term. Additionally, as noted, in a quantum system, the pairing between bra and ket states leads to a duplication of the cohomology, which can be identified with the antifields of the BV formalism. 

For $N=1$, we perform computations using multiple approaches: first through the partition function, then in the light-cone gauge, and finally via a detailed analysis of the covariant quantization and the equations of motion. We also consider an equivalent formulation in the large Hilbert space where cohomology is indeed trivial. 

Next, we recall the system of partial differential equations describing the cohomology, in the language of multiforms, reproduces the Dirac degrees of freedom. 
One may then wonder how the axial anomaly of the fermionic theory is represented in terms of p-forms. We observe that the axial anomaly is mapped to a Hodge anomaly for $p$-forms. This alteration does not affect the gauge invariance of the space-time theory. Instead, it implies that $p$- and $D-p$ forms are not dual at the quantum level. Understanding possible implications for geometry would be interesting, but we have not explored this here.  

Continuing with $N=1$, we discuss the emergence of the BV bracket in the operator formalism in different pictures and explicitly compute a quadratic action of the BF-type, which couples even and odd forms in target space of all degrees. In anology with $N=0$, we establish an equivalence between the field theory actions in ghost number zero and one with the latter represented by anti fields. We then then construct consistent interaction terms by means of a derived bracket based on the associative algebra of world line fields. This bracket introduces an interaction between even and odd forms, which is consistent on its own without requiring a quartic term. While this dynamical system is interesting in its own right, it would be even more intriguing to explore the corresponding interacting Lagrangian for Fermions using the correspondence described above.

In conclusion, this note presents several results that are absent from the existing literature, clarifying aspects of the geometrical derivation. We demonstrate that various geometric structures from string theory and theory of Riemann surfaces can be adapted to the spinning particle framework, including the construction of different pictures (where, in string theory, an infinite number of pictures exist) and the formulation of the target space action. The derivation of the Dirac action and its degrees of freedom appears to be naturally twisted in a geometric sense.


\section{N=0 particle} 
Although it is not the main focus of this paper, this section reviews some aspects of the bosonic $N=0$ particle. This serves as a warm-up for later, when a level of complexity is added with worldline supersymmetry.

\subsection{BRST analysis and target space action}
\label{secN0}
To set up the stage, we start with a particle model with no worldline supersymmetry.
The first-order Lagrangian for this model reads 
\begin{equation}
   S[x, P,e] = \int \, {d} \tau \,  \left( 
   P_{\mu} \dot{x}^{\mu} - e \frac{P^2}{2} \right)  \, . 
\end{equation}
where $e$ is an einbein implementing the constraint $P^2 = \eta^{\mu\nu} P_\mu P_\nu$. The associated BRST charge is 
\begin{eqnarray}\label{eq:Qc}
    Q = \frac12 c P^2 \, , 
\end{eqnarray}
where $c$ is a ghost field. After canonical quantization, our set of (anti-)commutators between operators is
\begin{eqnarray}
[x^{\mu}, P^{\nu}] = \eta^{\mu \nu}\,,  ~~~~ \{b, c\} =1 \, .
\end{eqnarray} 
$Q$ acts on the world line fields as follows:
\begin{eqnarray}
[Q, x^{\mu}] = c P^{\mu}\,, \qquad
[Q, P^{\mu}] =0\,, \qquad
[Q, c] =0\,, \qquad
[Q, b] = P^{2} \,.
\end{eqnarray}
We choose a vacuum (polarization) as 
\begin{eqnarray}
P^{\mu}|0\rangle =0\,, ~~~~~ b|0\rangle =0\,. 
\end{eqnarray}
Then the states of the Hilbert space $\mathcal{H}$ are constructed in terms of $x^{\mu}$ and $c$ as follows
\begin{eqnarray}
\label{haACA}
|\omega\rangle = (\phi_{0}(x) + c \phi_{1}(x) ) |0\rangle \, . 
\end{eqnarray}
To properly refine the partition function, we introduce a fugacity $q$ originating from the following scaling transformation of the fields 
\begin{eqnarray}
\label{scaleN=0}
x^{\mu} \rightarrow x^{\mu}\,, ~~~~~~
P^{\mu} \rightarrow q P^{\mu}\,,  ~~~~~~
c \rightarrow q^{-1} c\,,  ~~~~~~
b \rightarrow q^{2} b\,. 
\end{eqnarray}
Using this assignment, the partition function is  
\begin{equation}
\mathbb{P}(q,s) := \text{tr}_{\mathcal{H}}(s^{\mathbf{G}}q^{\mathbf{N}})
\end{equation}
where $\mathbf{N}$ is \eqref{scaleN=0} and we further refine with a ghost number fugacity $s$ 
\begin{eqnarray}
    \mathbb{P}(q,s) = (1 + s q^{-1}) \, .
    \label{part-N0}
\end{eqnarray}
This partition function implies that there are two cohomology classes with opposite parity and scaling dimension equal to -1. 

Now, we analyze $Q$-cohomology directly. On a generic state $\omega = \phi_0(x)|0 \rangle + c \phi_1(x) |0 \rangle$ the nilpotency of the BRST charge gives
\begin{eqnarray}
    Q \omega=0\,, ~~~~\Rightarrow \Box \phi_0(x) =0\,, 
\end{eqnarray}
so $Q$-closure puts no constraints on $\phi_{1}$. However $Q$-exactness $\delta \omega = Q \Lambda$, 
$\Lambda = \Lambda_0(x) + c \Lambda_1(x)$ yields 
\begin{eqnarray}
    \delta \phi_0(x) =0\,, ~~~~~ \delta \phi_1(x) = \Box \Lambda_0(x) \, .
\end{eqnarray}
Notice that we can use the gauge parameter $\Lambda_0(x)$ only if $\Box \phi_{1} \neq 0$. If $\Box \phi_{1}(x)= 0$, then we cannot use the gauge parameter $\Lambda_0(x)$ to remove $\phi_1(x)$. The final result is that the two cohomology classes $\phi_0(x)$ and $\phi_1(x)$ have different target ghost numbers ($0$ and $-1$), and both of them satisfy the Klein-Gordon equation. 
The partition function $\mathbb{P}(q,s)$ \eqref{part-N0} counts the two cohomology classes.

To construct a target space action, we define a BV anti-bracket as (the ground state is omitted) 
\begin{eqnarray}
\label{haA}
(\omega,\omega)_{BV} := \frac12 \int {\rm d}x^d {\rm d}c \, (\phi_{0} + c \phi_{1}) 
(\phi_{0} + c \phi_{1}) = \int {\rm d} x^d \, \phi_{0}(x) \phi_{1}(x)
\end{eqnarray}
therefore $\phi_{1}(x)$ has a BV interpretation of an antifield of $\phi_{0}$. We recall that the BRST transformation of $\phi_{1}$ is the equation of motion of $\phi_{0}$ by matching the BRST operator on the world line, $Q$, with the BV-BRST operator on the target space fields, $\mathfrak{s}$:
\begin{equation}
\label{haAA}
Q \omega = {\mathfrak s} \omega
\end{equation}
where
\begin{equation}
\label{haAB}
Q \omega = c \Box \phi_{0} = { \mathfrak s} \phi_{0} - c \, {\mathfrak s} \phi_{1}
\end{equation}
from which we get 
\begin{eqnarray}
\label{haAC}
{\mathfrak s} \phi_{0} =0\,, ~~~~~
{\mathfrak s} \phi_{1} = -  \Box \phi_{0}
\end{eqnarray}
that matches the usual BV framework for Klein-Gordon's theory.

The only sensible Hodge star operation  can be defined on the ghost $c$ as follows
\begin{eqnarray}
\label{haC}
\phi^* = \star (\phi_{0}  + c \phi_{1}) =   i \int e^{c \eta} 
(\phi_{0}  + \eta \phi_{1}) \, d\eta = i  ( - c \phi_{0}  + \phi_{1})
\end{eqnarray}
where $\eta$ is an auxiliary anticommuting variable. The $\star$ operation is idempotent. 

Therefore we obtain 
\begin{eqnarray}
\star \omega(c,x) = - i c \phi_0(x)|0 \rangle + i \phi_1(x) |0 \rangle \, ,
\end{eqnarray}
where the roles of $\phi_{0}$ and $\phi_{1}$ are interchanged. 
Note also that the pairing is invariant under the Hodge transformation: $(\omega, \omega)_{BV} = (\star \omega, \star \omega)_{BV}$. Finally, we can build the target space action (Klein-Gordon theory) with the data just discussed, as follows:
\begin{eqnarray}
\label{haD}
S[\phi] = \frac12 (\omega, Q\omega)_{BV} = \frac12 \int_{x,c}   (\phi_{0}  + c \phi_{1}) Q (\phi_{0}  + c \phi_{1})  = \frac12 \int d^{D}x \, \phi_{0} \Box \phi_{0} \, ,
\end{eqnarray}
which is the correct action of the field $\phi_{0}$. To add interactions to the action, we can introduce a generic function $f(\omega)$ and construct the new term 
\begin{eqnarray}
    \label{haDA}
    (\star \omega, f(\omega))_{BV} = \int
    (c \phi_0 + \phi_1) (f(\phi_0) +
    c \phi_1 f'(\phi_0)) = 
    \int d^Dx \Big(f(\phi_0) + \phi_0 f'(\phi_0) \Big) \, .
\end{eqnarray}
This gives a potential to $V(\phi_0) = f(\phi_0) + \phi_0 f'(\phi_0)$ for the physical degree of freedom $\phi_0$. For more discussion on interaction terms, see \cite{Doubek:2020rbg}

Inspection of cohomology shows that alongside the Klein-Gordon field $\phi_0$, there is also a second Klein-Gordon field $\phi_{1}$. Therefore, a natural question is whether one could write down an action functional for the dual field $\phi_{1}$ whose e.o.m.'s give the correct dynamics. This could work if the $\mathbb{Z}_2$ parity of $\phi_1$ (odd in the multiplet with $\phi_0$) is reversed to become even. Indeed, to obtain such action, we can use the Hodge star operator \eqref{haC} 
 \begin{eqnarray}
\label{haE}
S[\phi^{*}] = \frac12 (\star \omega, Q \star \omega)_{BV} = 
\frac12 \int_{x,c}  
(c\phi_{0}  + \phi_{1}) Q (c \phi_{0}  + \phi_{1})  = \frac12
\int d^{D}x \, \phi_{1}\Box \phi_{1} \, .
\end{eqnarray}
Now, the role of the field $\phi_{0}$ and of the antifield $\phi_{1}$ are interchanged. The above discussion is of course closely related to the two equivalent ground states of the $b,c$ ghost system in string theory.

\section{N=1 Particle}
In this section, we analyze the $N=1$ model. Following the previous sections, we compute the partition function and analyze the spectrum directly. Lastly, we construct the target space action and discuss an extension of this formalism in the large Hilbert Space.

\subsection{Worldline Fields, BRST and Partition Function}
The BRST charge of the $N=1$ particle model reads
\begin{eqnarray}
\label{aaA}
Q = c P^{2} + \gamma \psi\cdot P - \gamma^{2} b \, ,
\end{eqnarray}
where the worldsheet fields satisfy the following commutation relations: 
\begin{eqnarray}
\label{aaB}
[x^{\mu}, P^{\nu}] = \eta^{\mu \nu}\,, \quad 
[\beta, \gamma] = 1\,, \quad 
\{ \psi^{\mu}, \psi^{\nu} \} = \eta^{\mu\nu}\,, \quad
\{b, c\} =1 \, .
\end{eqnarray}
Explicitly, we have the action of $Q$ on the fields (we denote by $\mathbf{s}$ the BRST transformations as 
$[Q, \Phi] = \mathbf{s} \Phi$)
\begin{eqnarray}
\label{aaC}
\mathbf{s} x^{\mu} &=& c P^{\mu} + \gamma \psi^{\mu}\,,~~~
\mathbf{s} P^{\mu} = 0\,, ~~~~
\mathbf{s} \psi^{\mu} = \gamma P^{\mu} \,, ~~~~~
\mathbf{s} c = - \gamma^{2}\,, \nonumber \\ 
\mathbf{s} \gamma &=& 0\,, ~~~~~~~~~~~ 
\mathbf{s} \beta = \psi\cdot P -  \gamma b\,, ~~~~~~
\mathbf{s} b = P^{2} \, .
\end{eqnarray}
These BRST transformations are invariant under the rescaling 
\begin{eqnarray}
\label{aaD}
&&P^{\mu} \rightarrow q^{2} P^{\mu}\,, ~~~~~~
x^{\mu} \rightarrow x^{\mu}\,, ~~~~~
\psi^{\mu} \rightarrow q \psi^{\mu}\,, ~~~~ \nonumber \\
&&c \rightarrow q^{-2} c\,, ~~~~~
b \rightarrow q^{4} b\,, ~~~~
\gamma \rightarrow q^{-1} \gamma \,, ~~~~
\beta \rightarrow q^{3} \beta\,. 
\end{eqnarray}
Notice that $\hbar$ carries two dimensions. This is the reason why 
the difference between the scale of the field and its conjugate is two.\footnote{Note that 
on the Hilbert space, $P^{\mu} = \hbar \partial^{\mu}$ therefore the 
derivative does not carry any dimension since the dimension is carried by $\hbar$.} 

To compute the cohomology, we start by constructing the Fock space. We choose the following polarization: 
\begin{eqnarray}
\label{aaE}
P^{\mu} |0\rangle =0\,, ~~~~~
b |0\rangle =0\,, ~~~~~
\beta |0\rangle =0\,. 
\end{eqnarray}
For the fermions $\psi$, in even dimensions $D = 2 n$, we define two sets of fields $\psi_{i}, \bar\psi^{i}$ with 
$i=1,\dots, n$ and such that $\{\psi_{i}, \bar \psi^{j} \} = \delta^{j}_{i}$ and $\bar\psi^{i} |0\rangle =0$.  

In \cite{Boffo:2024lwd}, we described how the partition function counts the dimension of cohomology for the ${N}=2$ spinning particle. For ${N}=1$ we will consider the sum 
\begin{eqnarray}
\label{aaFA}
\mathbb{P}_{n}(q,{s}) := {\rm Tr}_{\mathcal{H}}[ s^{\mathbf{G}} q^{\mathbf{N}}] \,,
\end{eqnarray}
where $\mathbf{G}$ is the ghost number and 
$\mathbf{N}$ is the scale of the fields as discussed in \eqref{aaD}. 
Direct computation yields
\begin{eqnarray}
\label{aaF}
\mathbb{P}_{n}(q,s) = \frac{(1 + s q^{-2})(1 + q)^{n}}{(1 - s q^{-1})} \, ,
\end{eqnarray}
which counts the states obtained by $(\psi)^{f} c^{g} \gamma^{h} |0\rangle$ with 
$f=0,\dots, n$ and $g=0,1$, as well as $h\in \mathbb{N}^{*}$. Notice that $x^{\mu}$ does not scale 
and therefore, it does not contribute to the partition function. 

\subsection{Covariant Computation of the Partition Function}
\label{subsec-part-func}

The representation theory behind the $N=1$ model is that of Clifford algebra. To construct the spinorial representation thereof, one needs to choose a polarization of $\psi_i$'s, breaking the covariance under Lorenz symmetry. 

However, one can also do a covariant calculation of the partition function \eqref{aaFA}, as follows:
in order to properly take into account the degrees of freedom of $\psi$'s, 
we consider adding a new set of field $\psi'$ that removes $n$ degrees of freedom 
from the covariant set $\psi^{\mu}$. 
The new $n$ fields $\psi'$ are non-covariant and, therefore, need a new set of ghost-for-ghost $\psi''$, and this procedure of adding fields is iterated ad infinitum. 
To count the generation of the ghost-for-ghost, we introduce a new quantum number parametrized by $t$, which allows us to write the partition function (limited to the contribution from the $\psi$'s) as follows 
\begin{eqnarray}
\label{caA}
\mathbb{P}_{\psi}(q,t) = \prod_{p=0}^{\infty} (1 + t^{p} q)^{D(-1)^{p} }.
\end{eqnarray}
The exponent $(-1)^{p}$ is due to the alternating parity of the ghost-for-ghost sets. 
We are interested in the limit $t\rightarrow 1$, and for that we compute the log of $\mathbb{P}_{\psi}(q,t)$ as 
\begin{eqnarray}
\label{caB}
\lim_{t\rightarrow 1} \log \mathbb{P}_{\psi}(q,t) &=& D \lim_{t\rightarrow 1} 
\sum_{p=0}^{\infty} (-1)^{p} \log(1 +  t^{p} q) 
= 
D \lim_{t\rightarrow 1}  \sum_{r=0}^{\infty}  (\log(1 + t^{2r} q) -  \log(1 + t^{2r+1} q))
\nonumber \\
 && \hspace{-2cm} = D  \lim_{t\rightarrow 1}  \sum_{r=0}^{\infty}  
\sum_{l=0}^{\infty}\left( \frac{(-1)^{l}}{l} t^{2r l} q^{l}  -  \frac{(-1)^{l}}{l} t^{(2r+1) l} q^{l} \right)
=  D  \lim_{t\rightarrow 1} 
\sum_{l=0}^{\infty}   \frac{(-1)^{l}}{l}  q^{l}   \sum_{r=0}^{\infty}   \left( t^{2r l}-   t^{(2r+1) l} \right)
\nonumber \\
&& \hspace{-2cm} =  D  \lim_{t\rightarrow 1} 
\sum_{l=0}^{\infty}   
\frac{(-1)^{l}}{l}  q^{l} \left( \frac{1}{1 - t^{2l}} -  \frac{t^{l}}{1 - t^{2l}} \right) =  D  \lim_{t\rightarrow 1} 
\sum_{l=0}^{\infty}   
\frac{(-1)^{l}}{l}  q^{l} \frac{1}{1 + t^{l}} = \frac{D}{2} \log(1+q) .
\end{eqnarray}
From this, we get $\mathbb{P}_{\psi}(q,1) = (1+q)^{D/2}$, which is the same result obtained by using the non-covariant computation. 

\subsection{Two Dimensional Model}
\label{part-funct-2}

Let us consider the case $n=1, (D=2$) and set $s=-1$. Then, the partition function becomes  
\begin{eqnarray}
\label{aaG}
\mathbb{P}_{1}(q, -1) = -\frac1q + (1 - 1) + q \, .
\end{eqnarray}
The vanishing $q^{0}$ contribution is kept here for comparison with explicit expressions of the multiplet below. 
Indeed, let us consider lightcone fields. This corresponds to setting $P^0=P^1$ therefore, the relevant BRST transformations of our associative algebra of coordinates become
\begin{equation}
\label{abB}
\mathbf{s}\psi = 0, ~~~
\mathbf{s}\bar \psi =2\gamma P^0,~~~~
\mathbf{s} c = -\gamma^2, ~~~~
\mathbf{s} \gamma = 0.
\end{equation}
This clarifies which algebra elements are BRST invariant. In this polynomial ring, we find $\psi$ and $\gamma$, while $\gamma^2$ can be removed since this is BRST exact. Accordingly, in the lightcone frame  we have state solution of the cohomology is
\begin{eqnarray}
\label{lcgd2}
\omega_{lcf} = \left(  (\gamma+P^0c\psi) A^{(0)}_{1} + (A^{(0)}_{0} + \gamma \psi A^{(1)}_{1}) + \psi A^{(1)}_{0}\right) |0\rangle 
\end{eqnarray}
The numbers in superscript will later denote the form degree (the order of $\psi$), and those in subscript refer to the order of $\gamma$. 
{The first and last terms represent the cohomology in ghost number 1 and 0 respectively, with weight $q^{-1}$ and $q$.
$A^{(1)}_{1}$ is closed but also the image of $A^{(0)}_{0}$, thus contributing with opposite sign in $O(q^0)$ in \eqref{aaG}}. 

Now, we want to compare these results with a covariant computation. For that, we set the general expression (we omit writing the vacuum  $|0\rangle$ at the end of the 
expression, but it is supposed to be present)   
\begin{eqnarray}
\label{aaH}
 \nonumber 
\omega_{cov} = \sum_{p \in \mathbb{N}^{*}} \gamma^{p} \left(A^{(0)}_{p} + 
\psi^{\mu}  A^{(1)}_{p, \mu} + \frac12 \psi^2 A^{(2)}_{p}  \right)  + 
\sum_{p \in \mathbb{N}^{*}} c \gamma^{p} \left(\chi^{(0)}_{p} + 
 \psi^{\mu}  \chi^{(1)}_{p, \mu}+ \frac12 \psi^2 \chi^{(2)}_{p}  \right) 
\end{eqnarray}
where $\psi^{2} = \epsilon^{\mu\nu} \psi_{\mu} \psi_{\nu}$. 
Computing 
\begin{eqnarray}
    \label{legno3}
Q \omega_{cov} = 0 \,,     
\end{eqnarray}
we get the following equations for $p>  1$ 
\begin{eqnarray}
\label{aaI}
\chi^{(0)}_{p-1} &=& - \partial^{\mu} A^{(1)}_{p, \mu} \nonumber \\
\chi^{(1), \mu}_{p-1} &=& - \partial^{\mu} A^{(0)}_{p} - \epsilon^{\mu\nu} \partial_{\nu} A^{(2)}_{p}  \nonumber \\
\chi^{(2)}_{p-1} &=& - \epsilon^{\mu\nu} \partial_{\mu} A^{(1)}_{p, \nu}
\end{eqnarray}
 while for the cases $p=0$, the left-hand side vanishes. Therefore, we find non-trivial equations for $A_{0}^{(0)}, A^{(1)}_{0,\mu}$ and $A^{(2)}_{0,\mu\nu}$ whereas for $p\geq1$ one can solve the equations for $\chi_{p}$'s.
 
 Then, substituting such solutions into $\omega_{cov}$ 
 we get
 \begin{eqnarray}
\label{aaL} \nonumber 
\omega_{cov}  &=& 
\sum_{p = 0,1} \gamma^{p} \left(A^{(0)}_{p} + 
\psi^{\mu}  A^{(1)}_{p, \mu} + \frac12 \psi^2 A^{(2)}_{p}  \right)  
+ c (d+d^{\dagger}) \left(A^{(0)}_{1} + 
\psi^{\mu}  A^{(1)}_{1, \mu} + \frac12 \psi^2 A^{(2)}_{1}  \right)
\nonumber \\ 
&+& 
Q \left[\sum_{p\geq 1}  c \gamma^{p-1} \left( A^{(0)}_{p+1} + 
\psi^{\mu}  A^{(1)}_{p+1, \mu} + \frac12 \psi^2 A^{(2)}_{p+1}  \right) \right] 
\end{eqnarray}
therefore, apparently, only the $(p=0)$-terms are in the cohomology. 
The $(p=1)$-terms cannot be proportional to $c \gamma^{p-1}$, 
but we can introduce a $c$-independent expression in the $Q$-exact terms as 
$Q \bf{C}_{0}$ -- where ${\bf C}_{0} =  \left(C^{(0)}_{0} + 
\psi^{\mu}  C^{(1)}_{1, \mu} + \frac12 \psi^2 C^{(2)}_{1}  \right)$ is a multiform --
to remove the $\mathbf{A}_{1} = \left(A^{(0)}_{1} + 
\psi^{\mu}  A^{(1)}_{1, \mu} + \frac12 \psi^2 A^{(2)}_{1}  \right)
$ terms from the cohomology by the gauge transformations 
\begin{eqnarray}
\label{aaLA}
\delta {\mathbf A}_{1} = (d+d^{\dagger}) \bf{C}_{0}\,,
\end{eqnarray}
 unless that $ {\mathbf A}_{1}$ were in the kernel of $(d+d^{\dagger})$, similarly to what observed 
 in the previous section for the scalar field $\phi_1$. In that case, the term proportional to $c$ in 
 \eqref{aaL} drops out. Therefore, we have two sets of 
 cohomology classes, one for the multiforms $ {\mathbf A}_{0}$ and one for the multiforms $ {\mathbf A}_{1}$. 
 The terms in the square brackets are BRST exact and do not contribute to the cohomology. 
This is consistent with the partition function \eqref{aaG} where $q^f$-terms
vanish for $f>1$.  

The equations for the non-exact terms $(p=0,1)$ are 
\begin{eqnarray}
\label{aaI}
0 &=& \partial^{\mu} A^{(1)}_{p, \mu} \nonumber \\
0 &=& \partial^{\mu} A^{(0)}_{p} + \epsilon^{\mu\nu} \partial_{\nu} A^{(2)}_{p}  \nonumber \\
0 &=& \epsilon^{\mu\nu} \partial_{\mu} A^{(1)}_{p, \nu} \, .
\end{eqnarray}
Notice that acting with $\partial^{\mu}$, these equations imply the Klein-Gordon 
equations 
\begin{eqnarray}
\label{aaI}
\partial^{2} A^{(1)}_{p, \mu}  = 0\,, ~~~~~
\partial^{2} A^{(0)}_{p} = 0\,, ~~~~~~
\partial^{2} A^{(2)}_{p} =0 \, .
\end{eqnarray}
To compare with \eqref{lcgd2}, it is convenient to set $P_{\mu} = (p_{+}, \partial_{-})$ where the first component $P_{+}$
is set to a constant value $p_{+}$ while the second component $P_{-}$ is identified with the differential 
operator $P_{-} = \partial_{-}$. Then, the equations become 
(setting $A^{(1)}_{p,\mu} = (A^{(1)}_{p,+}, A^{(1)}_{p,-})$) 
\begin{eqnarray}
\label{aaJ}
{p_+(A^{(2)}_p -A^{(0)}_p)=0,} 
&&~~~~~~~
{\partial_-(A^{(2)}_p + A^{(0)}_p)=0 ,} \nonumber \\
p_{+} A^{(1)}_{p,-} -  \partial_{-} A^{(1)}_{p,+} =0\,, && ~~~~~~
p_{+} A^{(1)}_{p,-} +  \partial_{-} A^{(1)}_{p,+} =0\,. ~~~~~~
\end{eqnarray}
We can solve the equations in the first column by
\begin{eqnarray}
\label{aaK}
{A^{(2)}_{p}  =  A^{(0)}_{p}}\, ,\hspace{1cm}
A^{(1)}_{p,-} = \frac{1}{p_{+}} \partial_{-} A^{(1)}_{p,+} \,,
\end{eqnarray}
valid for $p=0,1$. Notice that inserting these solutions 
in the other equations, one recovers the Klein-Gordon equations 
\begin{eqnarray}
\label{aaM}
 {\eta^{+-} p_+\partial_- A^{(1)}_{p, +} =0, \qquad  \eta^{+-} p_+\partial_- A^{(0)}_p =0\, .}
\end{eqnarray} 
Therefore, we see that in solving the equations of motion, we can compare with the components 
given in $\omega_{lcf}$ in \eqref{aaG}. We are left with the independent components $A^{(0)}_{p}, A^{(1)}_{p,+}$ with $p=0$. The ghost number and $q$-weights of these states match those with those appearing in the partition function. However unlike \eqref{lcgd2} they are all in the cohomology in the covariant formulation.  This apparent o/contradiction is resolved if we recall that \eqref{aaG} has the interpretation of the Euler characteristics which counts the weighted sum over states. In the light cone frame the 2-form was not included in the Hilbert space. In the covariant formulation it is present and promotes $A_1^{(0)}$ to a physical state. 


\subsection{Action}
We would now like to construct the action by reproducing the above cohomology from its equations of motion modulo gauge redundancies. We first define the BV symplectic structure using a suitable measure on the worldline fields. Notice that the Hilbert space contains the fields $x^\mu, \psi^\mu, c$ and $\gamma$. We integrate over the 
fermionic fields $c, \psi^\mu$ with a Berezin integration formula, the integration over $x^\mu$ is the conventional Riemann-Lebesgue integral, but for $\gamma$ some care is needed. Indeed, a naive integration would lead to a divergent result. Therefore, it is customary to introduce a special measure to regulate this integral. This is done by introducing a Picture Changing Operator (PCO), which is a picture number $+1$ element of the cohomology, and a choice of a convenient representative is given by 
\begin{eqnarray}
    \label{pcA}
    {Y}(c,\gamma) = c \delta'(\gamma) \, .
\end{eqnarray}
where $\delta'(\gamma)$ is the derivative of the Dirac delta distribution  \cite{Belopolsky:1997bg}. 
Finally, we can write the BV symplectic structure as follows 
\begin{eqnarray}
\label{aaMA}
(\omega_{cov}, \omega_{cov})_{BV}= \frac12
\int_{x,c,\gamma,\psi} \omega_{cov}\wedge \omega_{cov} Y(c,\gamma).
\end{eqnarray}
A simple calculation gives 
 \begin{eqnarray}
\label{aaMB}\nonumber 
(\omega_{cov}, \omega_{cov})_{BV} 
 = \int_{_{x,c,\gamma,\psi} } \hspace{-.7cm}
 {\bf A}_{1}  {\bf A}_{0}  \gamma c \delta'(\gamma) 
=\int d^Dx d^D\psi {\bf A}_{0}  {\bf A}_{1} \, .
\end{eqnarray}
Computing the Berezing integral over $\psi$'s leads to the pairing between fields and antifields. In fact, as for $N=0$, 
imposes that the role of antifield is played by the (multi)forms in the second cohomology class.


Now, we are ready to compute the action for the kinetic term. 
\begin{eqnarray}
\label{aaZC}
S[{\bf A}_{0}] &=&\frac12 \int_{_{x, c, \gamma, \psi}} \omega_{cov} \wedge Q \omega_{cov} \, Y(c,\gamma) \nonumber \\
&=& \frac12 \int_{_{x, c, \gamma, \psi}}\sum_{p\geq0} (\gamma^{p} A_{p} + c \gamma^{p} \chi_{p}) \ Q 
\sum_{p'\geq0} (\gamma^{p'} A_{p'} + c \gamma^{p'} \chi_{p'}) c\delta'(\gamma)  \nonumber \\
&=& 
 \frac12 \int_{_{x, c, \gamma, \psi}}\sum_{p\geq0} (\gamma^{p} A_{p} ) \left(\gamma (d+d^{\dagger}) \right)
\sum_{p'\geq0} (\gamma^{p'} A_{p'}) c\delta'(\gamma)  \nonumber \\
&=& 
 \frac12 \int_{_{x, c, \gamma, \psi}} {\bf A}_{0} \left(\gamma (d+d^{\dagger}) \right)
{\bf A}_{0} c\delta'(\gamma)  =  \frac12 \int d^{D}x d^{D}\psi \, 
{\bf A}_{0} \left(d+d^{\dagger}\right) {\bf A}_{0}  
\end{eqnarray}
which is the Dirac Lagrangian for the multiform ${\bf A}_{0}$. Note that the $c P^{2}$ 
term of the BRST charge drops out thanks to the ghost $c$ in the PCO $Y(c, \gamma)$. The derivative on the $\delta(\gamma)$, by integration 
by parts, soak up the $\gamma$ in the $\gamma (d+d^{\dagger})$ term of the BRST charge $Q$. 
The last term $-\gamma^{2} b$ drops out because of the $\gamma^{2}$ term which cancels because of $\delta'(\gamma)$. 
Note that all other terms (of the trivial part of the cohomology) drop out, and only the ${\bf A}_{0}$ part is saved. This is reminiscent of the Maurer-Cartan action for Chern-Simons in $D=4$ of \cite{Jurco:2018sby}. 
\vskip .5cm 
An important remark is that the action corresponds to the Dirac action for a multiform field, confirming that the physical states of the present sector are indeed Dirac fermions. The differential operator $d + d^\dagger$ squares to $\square$. 
Another remark is: as in the case of $N=0$, we can wonder whether it is possible to construct an action describing the second multiform ${\bf A}_1$ representing the second copy of the cohomology discussed above. For that, we need a preliminary discussion on pictures and PCOs in the present framework. 

\subsection{Picture One}

As in \cite{Boffo:2024lwd}, we can have copies of the same  cohomology through the  different pictures. In the present framework, as it has already been advocated, the picture counts the number of $\delta(\gamma)$, and since this symbol has Grassmann parity, we can admit only one picture in the game. Together with the notion of the picture, we have Picture Changing operators (as seen on eq.\eqref{pcA}) $Y(c, \gamma)$, raising the picture number, and $Z(b, \beta)$, lowering the picture. Furthermore, there is another way to change the picture by computing the Hodge dual of the vertex $\omega$. Indeed, by supergeometric considerations, the Poincar\'e dual $\omega$ is in the picture one sector. This is defined as follows 
\begin{eqnarray}
\label{aaZZA}
\star \omega(x^\mu, \gamma, c, \psi^\mu) = \# \int e^{i(\sigma \gamma + \eta c + \rho_{\mu} 
\eta^{\mu\nu} \psi_{\nu})} \omega(x^\mu,\sigma, \eta^\mu, \rho)  [d\eta d^{D}\rho d\sigma] 
\end{eqnarray}
where $\sigma$ are commuting variables and $\eta, \rho_{\mu}$ are anticommuting variables. 
The integration ``measure'' $[d\eta d^{D}\rho d\sigma]$ reminds us of the integration variables. The factor $\#$ is needed to impose idempotency and will be neglected in the following. $\eta^{\mu\nu}$ is a flat metric on the spacetime manifold. 

Computing the Hodge dual of the covariant expression -but with the $Q$-exact terms left out of our considerations-, letting $*$ be the spacetime Hodge star ($*\psi^\mu \sim * dx^\mu = \epsilon^{\mu \mu_2 \dots \mu_D} \eta_{\mu_2 \nu_2} \dots \eta_{\mu_D \nu_D}  \\ dx^{\nu_2} \wedge \dots \wedge dx^{\nu_D}$), we get 
\begin{eqnarray}
\label{aaZA}
\star \omega_{cov} &=& \star 
\left( {\bf A}_{0} +  \gamma {\bf A}_{1} \right) = 
 c\left( *{\bf A}_{0} \delta(\gamma) + *{\bf A}_{1}\delta'(\gamma) \right)
\end{eqnarray}
to be compared with the picture-raised vertex (obtained by applying \eqref{pcA})
\begin{eqnarray}
\label{aaZAB}
Y(c, \gamma)\omega_{cov} = 
 c\left( {\bf A}_{0} \delta'(\gamma)  - {\bf A}_{1}\delta(\gamma) \right) \, .
\end{eqnarray}
Therefore, it becomes possible to impose a self-duality condition:
\begin{eqnarray}
    \label{legnoA}
Y \omega_{cov} = \star \omega_{cov}   . 
\end{eqnarray}
We will return to this possible Hilbert space restriction at the end of this section. 

In sec. \ref{secN0} we found the (Klein-Gordon) action for the second copy of the cohomology. Therefore, we would like to do the same here: first of all, we show that the action of the BRST transformation on the Hodge dual vertex $\star \omega_{cov}$ leads to the same equations of motion, and then we provide the corresponding action. 
Acting with $Q$ on $\star \omega_{cov}$, we 
have 
\begin{eqnarray}
    \label{legnoB}
    Q \star \omega_{cov} = \delta(\gamma) c (d+ d^\dagger) * {\bf A}_1 =0 \, ,
\end{eqnarray}
leading to the same differential equations as in the picture zero case. 
However, we have to underline that in eq. \ref{legno3} we find an equation for the multiform ${\bf A_0}$, and the second equation is obtained by imposing the equivalence up to $Q$-exact terms. Here, the role is interchanged, we get an equation for 
${\bf A_1}$ and the equation for ${\bf A_0}$ is obtained by requiring the non-exactness of the results.  (This is analogous to self-duality for the $2$-form field strength $F = * F$ in $D=4$, where the closure of $F$ is a consequence of Bianchi identities and that of  $* F$ is due to equations of motion.) 
By this fact, we use the Hodge dual field $\star \omega_{cov}$ to construct the corresponding action. 
However, in contrast with the bosonic sector in $N=0$, the Hodge dual of 
$\omega_{cov}$ has picture one. This implies that the product $\star \omega_{cov}$ with another picture one 
vertex vanishes. To avoid this, we have to reduce the picture by using the PCO $Z$
\begin{eqnarray}
\label{aaZD}
Z (b,\beta) = [Q, \theta(\beta)] = \psi\cdot P \delta(\beta) -  b \delta'(\beta)
\end{eqnarray}
where $\theta(\beta)$ is the Heaviside distribution; it is easy to check that $[Z(b,\beta), Y(c, \gamma)] =1$. 
Acting on the Hilbert space,  $\beta$ and $P^{\mu}$ are replaced by 
the differential operators $\partial_{\gamma}, \partial_{\mu}$. Since $Z(b,\beta)$ is BRST invariant 
we can insert it where it is more convenient; therefore, we set 
\begin{eqnarray}
\label{aaZE}
S[{\bf A}_{1}] &=& \frac12 \int_{_{x,c,\gamma,\psi}} \left(  (Z(b, \beta) \star \omega_{cov})  Q \star \omega_{cov} \right) \nonumber \\
&=&\frac12  \int d^{D}x d^{D} \psi \, (\ast{\bf A}_{1}) (d + d^{\dagger}) (\ast {\bf A}_{1}) = \frac12
\int d^{D}x d^{D}\psi \,{\bf A}_{1} (d + d^{\dagger}) {\bf A}_{1} 
\end{eqnarray}
which is again the Dirac Lagrangian for the multiform ${\bf A}_{1}$.\footnote{It is obvious that 
$\star (d + d^{\dagger}) \star = (d^{\dagger} + d)$ and the plus in the combination is crucial.  
} This confirms the results of the cohomology. 
Note the duality between the fields ${\bf A}_{0}$ and the antifields ${\bf A}_{1}$ by Hodge duality. 
\vskip .5cm 

Finally, if we impose the condition that the 
vertex should be self-dual as in \ref{legnoA}, then we have 
\begin{eqnarray}
\label{sdC}
S[{\bf A}_{1}] &=& \frac12 \int_{_{x,c,\gamma,\psi}} \left(  (Z(b,\beta) \star \omega_{cov})  Q \star \omega_{cov} \right) \nonumber 
\\
&=& \frac12 \int_{_{x,c,\gamma,\psi}} \left(  (Z(b,\beta) Y(c,\gamma) \omega_{cov})  
 Q Y(c,\gamma)\omega_{cov} \right) \nonumber \\ 
 &=& 
 \frac12 \int_{_{x,c,\gamma,\psi}} \left(  Y(c,\gamma) Z(b,\beta) Y(c,\gamma) \omega_{cov}  
 Q \omega_{cov} \right) \nonumber \\ &=& 
 \frac12 \int_{_{x,c,\gamma,\psi}} \left(  Y(c,\gamma)\omega_{cov}  
 Q \omega_{cov} \right) = S[{\bf A}_0]
 \end{eqnarray}
where we used the property $ Y(c,\gamma) Z(\beta) Y(c,\gamma) =  Y(c,\gamma)$ valid 
for any choice of PCO and the property that $[Q,  Y(c,\gamma)] = 0$. With the same identities, one can prove that also the anti-bracket is preserved by the Hodge duality. 
Eq. \eqref{sdC} shows the equivalence between the action for ${\bf A}_0$ and ${\bf A}_1$ and, therefore, the self-duality of it.

We conclude this section by cross-checking our statement about the existence of another copy of the cohomology at picture one, again by using the partition function to study the dual expression. We first consider the Hilbert space with the following states $(\psi)^{f} c^{g} \delta^{(h)}(\gamma) |0\rangle$ with $f=0,\dots,n$ and $g=0,1$ while $ h \in 
\mathbb{N}^{*}$. Notice that $\delta(\gamma)$ scales with the dimension $q$ and its $h$-derivatives scale as $q^{h+1}$. Then, the partition function in picture 1 is 
\begin{eqnarray}
\label{aaZ}
\mathbb{P}^{(1)}_{n} (q, -1)= \frac{(1 - q^{-2})(1+q)^{n}}{(1+q)} q = - \frac1q (1-q)(1+q)^{n}
\end{eqnarray}
The factor $q$ corresponds to $\delta(\gamma)$ while the factor $(1+q)^{-1}$ corresponds to derivatives of the delta function $\delta(\gamma)$.\footnote{A derivative with respect to $\gamma$, namely $\partial_{\gamma} \delta(\gamma)$ scales as $q$. Note that $\beta$ acts on the Hilbert space as the differential operator $\partial_{\gamma}$ but scales as $q^{3}$. The mismatch is due to the $\hbar$ contained in $\beta$ as discussed above.}

Again, computing the Hodge dual of the lightcone frame expression, we get 
\begin{eqnarray}
\label{aaZA}
\star \omega_{lcf} &=& \star 
\left( (\gamma + c P^0 \psi) A^{(0)}_{1} + (A^{(0)}_{0} + \gamma \psi A^{(1)}_{1}) + \psi A^{(1)}_{0}\right) = 
\nonumber \\
&=& \left(c\psi \partial_\gamma + P^0\right)A_{1}^{(0)} \delta(\gamma) + c\left((\psi \delta(\gamma) A^{(0)}_{0} + 
\delta'(\gamma) A^{(1)}_{1}) +  \delta(\gamma) A^{(1)}_{0} \right) 
\end{eqnarray}
This matches perfectly, upon a rescaling by $q^{-1}$, with the partition function for picture zero: $\frac{1}{q} \mathbb{P}_{1}(q,-1) = \frac{1}{q}(q  + (1-1) - q^{-1})$. 



\subsection{Four dimensional model} 
\label{part-func-4}
In the four dimensional case, the number of dof's changes, but the results of the previous section remain the same. In particular, the formula for the action, PCO, the Hodge dual expression $\star \omega$ and the relations between these ingredients are the same. On the contrary, we have more dof's coming from the fact that the multiforms ${\bf A}_0$ and ${\bf A}_1$ contain up to $4$-forms. Notwithstanding, there is crucial difference: in $D=4$ the Dirac equation halves the number of dof's. Then, we have half of the dof's because of the equations of motion. We will furthermore identify an infinite gauge symmetry which allows us to halve the degrees of freedom, and we get the expected number to be compared with the partition function. However in picture zero these are not BRST transformations and one may thus wonder why the result should agree with the partition function. The resolution comes by noting that these symmetries do become BRST after inserting a picture changing operator needed for a well defined pairing. The partition function appears to be "aware" of this fact. 

The partition function is 
\begin{eqnarray}
\label{aaN}
\mathbb{P}_{2}(q,-1) = -\frac1q + (1-2)  + (2 - 1) q + q^{2} = -\frac1q - 1  + q + q^{2} \, .
\end{eqnarray}
Again, we leave the differences $(1-2)$ and $ (2 - 1) q$ in order to compare with explicit expressions of the states in cohomology. As for $n=1$, the light-cone gauge expression lists all independent fields:
\begin{eqnarray}
\label{aaP}
\omega_{lcf} &=& (\gamma + c \psi^+ P^+) A^{(0)}_{1} + (A^{(0)}_{0} + (\gamma + c \psi^+ P^+) \psi^{i} A^{(1)}_{1,i}) + \nonumber \\
&+&(\psi^{i}  A^{(1)}_{0,i} + \gamma \psi^{+} \psi^{t} A^{(2)}_{1, +t}) + \psi^{+} \psi^{t} A^{(2)}_{0, +t}, 
\end{eqnarray}
where we have deployed $+$ and $t$ (for transverse) to denote the non-zero components and coordinates and $i=+,t$. The second and the third group of terms (inside the brackets) correspond to the contribution $(1-2)$ and $(2 - 1) q$. Moreover, the series after the last equality should be compared to the lightcone expression:
\begin{align}
   \tilde\omega_{lcf} = A_+\psi^++A_{+ t}\psi^+\psi^t + (\gamma + c \psi^+ P^+) A + (\gamma + c\psi^+P^+)\psi^t A_t.
\end{align}

Let us check the covariant equations. Again, only the contributions for $p=0,1$ correspond to cohomologies. 
Because of the same argument that we used in the discussion of the $2$-dimensional case, we can construct the covariant expressions as follows: 
\begin{eqnarray}
\label{aaQ}
\omega_{cov} &=& \left( A^{(0)}_{0} + \psi^{\mu}A^{(1)}_{0, \mu} + \frac12 \psi^{\mu} \psi^{\nu} 
A^{(2)}_{0, \mu\nu} +  \frac16 \psi^{\mu} \psi^{\nu} \psi^{\rho} A^{(3)}_{0, \mu\nu\rho} 
+  \frac{1}{24} \psi^{\mu} \psi^{\nu} \psi^{\rho} \psi^{\sigma }A^{(4)}_{0, \mu\nu\rho\sigma} \right) \nonumber \\
\hspace{-.5cm}&+& \gamma \left( A^{(0)}_{1} + \psi^{\mu}A^{(1)}_{1, \mu} + \frac12 \psi^{\mu} \psi^{\nu} 
A^{(2)}_{1, \mu\nu} +  \frac16 \psi^{\mu} \psi^{\nu} \psi^{\rho} A^{(3)}_{1, \mu\nu\rho} 
+  \frac{1}{24} \psi^{\mu} \psi^{\nu} \psi^{\rho} \psi^{\sigma }A^{(4)}_{1, \mu\nu\rho\sigma} \right),
\end{eqnarray}
which satisfy the covariant equations
\begin{eqnarray}
\label{aaR}
&&\partial_{\mu} A^{(0)}_{p} + \partial^{\nu} A^{(2)}_{p, \mu\nu} =0\,, ~~~~~\nonumber \\
&&\partial_{[\mu} A^{(1)}_{p, \nu]} + \partial^{\rho} A^{(3)}_{p, \mu\nu\rho} =0\,, ~~~~\nonumber \\
&&\partial_{[\mu} A^{(2)}_{p, \nu \rho]} + \partial^{\sigma} A^{(4)}_{p, \mu\nu\rho\sigma} =0\,, ~~~~~ \nonumber \\
&&\partial^{\mu} A^{(1)}_{p, \mu} =0\,, ~~~~~~\nonumber \\
&&\partial_{[\mu} A^{(3)}_{p, \nu\rho\sigma]} =0\,. 
\end{eqnarray}

In terms of the multiforms, the covariant equations \eqref{aaR} can be written as:
\begin{equation}
    ({d} + {d}^\dagger) {\bf A}_p = 0 \, .
\end{equation}
At this point it is crucial to observe that the equations have an infinite tower of gauge symmetries: 
\[\delta {\bf A}_p =( {d} + {d}^\dagger )\lambda_{p,0}, \, \delta \lambda_{p,0} =( {d} + {d}^\dagger )\lambda_{p,1}\] 
and so on. 
Hence if $d$ refers to the d.o.f.'s, we are counting 
\[
d \, \sum_{n=0}^{+\infty} (-1)^n = d\,  \underset{q\rightarrow 1}{\text{lim}} \,  \sum_{n=0}^{+\infty} (-q)^n = d\, \underset{q\rightarrow 1}{\text{lim}}  \frac{1}{1+q} = \frac{d}{2} \, .
\]
Since the number of d.o.f.'s is $d = 4$, we are left with 2. 

We would now like to provide more details on how to solve the equations \eqref{aaR}. In the notation $A^{(k)}$, that leaves the subscript $p=0,1$ implicit ($\, k=0, \dots, 3$), let us denote the spacetime indices of the forms as $(0,i,3), i=1,2$. Possible confusion between the form degree and the component of the form should be avoided by the extra round brackets on the former. We fix the metric to be Minkowski with signature $(+,-,-,-)$. Then it is possible to rearrange the set of differential equations for $A^{(k)}$, with $k$ even, in the following fashion:
\begin{align}
& \begin{pmatrix}
\partial_0 & \partial_3 \\
\partial_3 & -\partial_0
\end{pmatrix} \begin{pmatrix}
A^{(0)} \\ A^{(2)}_{03}
\end{pmatrix} =  \partial^j \begin{pmatrix} 
A^{(2)}_{0j} \\ A^{(2)}_{3j}
\end{pmatrix}\, , &  &\begin{pmatrix}
\partial_0 & \partial_3 \\
\partial_3 & -\partial_0
\end{pmatrix} \begin{pmatrix}
A^{(2)}_{ij} \\ A^{(4)}_{0ij3}
\end{pmatrix} = \partial_{[i} \begin{pmatrix} 
A^{(2)}_{0 \vert j]} \\ A^{(2)}_{3 \vert j]}
\end{pmatrix} \, ,\\
& \begin{pmatrix}
\bm{1}_2 \partial_0 & \bm{1}_2 \partial_3 \\
\bm{1}_2\partial_3 & -\bm{1}_2 \partial_0
\end{pmatrix} \begin{pmatrix}
A^{(2)}_{i0} \\ A^{(2)}_{i3}
\end{pmatrix} =  \begin{pmatrix} 
\partial^k A^{(2)}_{ik} - \partial_i A^{(0)} \\ \partial^j A^{(4)}_{0ij3} -\partial_i A^{(2)}_{03}  
\end{pmatrix} \, . & & 
\label{eq-row2}
\end{align}
All in all, these show that we can express all the components in terms of the fields $A^{(2)}_{i0}$ and $A^{(2)}_{i3}$, which sum up to 4. Then the presence of an infinite tower of gauge symmetries halves this number to 2.

We cannot yet compare with the counting from the partition function, as we shall still analyse the odd forms. Similar arguments will help us conclude that the propagating d.o.f.'s in cohomology are 2. Regarding the differential equations for $(A^{(1)},A^{(3)})$ we can make the following arrangement:
\begin{align}
& \begin{pmatrix}
\bm{1}_2 \partial_0 & \bm{1}_2 \partial_3 \\
\bm{1}_2\partial_3 & -\bm{1}_2 \partial_0
\end{pmatrix} \begin{pmatrix}
A^{(1)}_{i} \\ A^{(3)}_{0i3}
\end{pmatrix} =  \begin{pmatrix} 
\partial_i A^{(1)}_{0} + \partial^j A^{(3)}_{0ij} \\ \partial_i A^{(1)}_3 + \partial^j A^{(3)}_{3ij}
\end{pmatrix} \, , & & \\
 & \begin{pmatrix}
 \partial_0 & \partial_3 \\
\partial_3 &  -\partial_0
\end{pmatrix} \begin{pmatrix}
A^{(1)}_0 \\ A^{(1)}_3
\end{pmatrix} =  \partial^i \begin{pmatrix} 
 A^{(1)}_{i} \\  A^{(3)}_{0i3}
\end{pmatrix} \, , & \; & \begin{pmatrix}
 \partial_0 & \partial_3 \\
\partial_3 &  -\partial_0
\end{pmatrix} \begin{pmatrix}
A^{(3)}_{ij0} \\ A^{(1)}_{ij3}
\end{pmatrix} = \partial_{[i} \begin{pmatrix}
    A^{(1)}_{j]} \\ A^{(3)}_{0\vert j] 3}
\end{pmatrix}.
\label{diverg 1-3}
\end{align}
Therefore we are able to express all of the odd forms as functions only of $A^{(3)}_{0i3} , \, A^{(1)}_i$, which has four components in total. 
Then the residual gauge for gauge symmetries fixes the number of components to be 2.

\vskip .5cm
As for the odd-symplectic form, we can let that be
\begin{eqnarray}
\label{aaWA}
(\omega_{lcf}, \omega_{lcf}) &=& \int_{_{x,c,\gamma, \psi_{i}}}
\hspace{-.7cm}
\omega_{lcf}\wedge \omega_{lcf} Y(c,\gamma) \nonumber \\
&=& 
\int_{_{x,c,\gamma, \psi_{i}}}
\hspace{-.7cm} \left[\left(A_{1}^{(0)} A^{(2)}_{0, ij} + A_{0}^{(0)} A^{(2)}_{1, ij} + A^{(1)}_{0,i} 
A^{(1)}_{1,j}\right) \gamma \psi^{i} \psi^{j} \right] c\delta'(\gamma) \nonumber \\
&=&
\int d^{4}x 
\left(A_{1}^{(0)}A^{(2)}_{0, ij} + A_{0}^{(0)} A^{(2)}_{1, ij} + A^{(1)}_{0,i} 
A^{(1)}_{1,j}\right)  \epsilon^{ij}  \, .
\end{eqnarray}
Once again, we see a perfect match between the fields and antifields.  
The action for both ${\bf A}_0$ and ${\bf A}_1$ is obtained exactly in the same way as for the previous case. The only relevant difference is the Berezin integration over $\psi$.

The "behind-the-scenes" correspondence with fermionic fields, that will be more thoroughly presented in Sec. \ref{correspondence}, should convince that there are several ways to built interactions. 
In the next to the following section, we present an original method, based on a derived bracket known as BV bracket (which is an example of Gerstenhaber algebra). The BV bracket is constructed with the BRST operator and we will exploit it for the construction of new interactions between four fermions. Before that, let us pause for a self-contained discussion about the Large Hilbert Space.

\subsection{Large Hilbert Space}
In this section we would like to embed the previous construction into an analog of Large Hilbert Space known from SFT literature \cite{Erler:2013xta}. 
In the Large Hilbert Space one works with the operators 
\begin{eqnarray}
\label{lhsA}
\frac1\gamma\,, \hspace{.5cm}
\xi\equiv \theta(\beta)\,,  \hspace{.5cm}
\eta = \lim_{\epsilon \rightarrow 0} \sin[\epsilon \beta]
\end{eqnarray}
satisfying the following algebra (in the sense of distributions):
\begin{eqnarray}
&&\eta \gamma^{-p}  = (-1)^{p} \delta^{(p-1)}(\gamma)\,, \hspace{.7cm}
\theta(\beta) \delta^{(p)}(\gamma) = (-1)^{p} p! \gamma^{-p-1} \, ; \,, \hspace{.7cm} \label{lhsC} \nonumber \\
 &&\{\eta, \theta(\beta)\}  = 1\,, \hspace{2.2cm}
[Q, \eta] =0 \, . \label{lhsB}
\end{eqnarray}
All of the above relations can be verified in the distributional sense. For instance $[Q,\eta]=0$ can be proven by using the distributional limit $\epsilon \rightarrow0$. The Large Hilbert Space (LHS) is an enlargement of the Small Hilbert Space (SHS) with the states generated by the inverse of $\gamma$ as well as those generated by the Heaviside $theta$ function $\theta(\beta)$ and its derivatives. 
Therefore, the space of distributions is enlarged and in such a new framework some simplifications occur. 
The operator $\eta$ plays the role of a new BRST differential needed to consistently bring the LHS to the SHS, as shown by the first equation in \eqref{lhsC}. 
For more, see also \cite{Catenacci:2018xsv}. 

We shall now show that in the Large Hilbert Space the $Q$-cohomology is trivial. First notice that 
 \begin{eqnarray}
\label{lhsC}
\Big[Q, \frac{c}{\gamma^{2}}\Big]= 1\,, \hspace{.7cm}
\Big[\eta,  \frac{c}{\gamma^{2}}\Big] = Y(c, \gamma) \, .
\end{eqnarray}
Then, the $N=1$ BRST charge can be shown to be given by conjugation on $-\gamma^2 b$ as 
\begin{eqnarray}
\label{lhsD}
Q = e^{\gamma^{-1} c \psi\cdot P} (- \gamma^{2} \partial_{c})  e^{- \gamma^{-1} c \psi\cdot P}
\end{eqnarray}
where the operator $\gamma^{-1} c \psi\cdot P$ is defined only in the Large Hilbert Space. 

Alternatively, one can see that the cohomology is trivial from direct calculation. 
Going to the Large Hilbert Space implies that the states are defined as follows 
\begin{eqnarray}
\label{lhsDA}
\omega_{LHS} = \sum_{p=-\infty}^{\infty} \gamma^{p} A_{p} \ket{0} + c \gamma^{p} \chi_{p} \ket{0}, \quad \xi \ket{0} = 0 
\end{eqnarray}
where, in contrast with the SHS, here the summation is extended form $-\infty$ to $\infty$. 
Then for the kernel of the BRST charge, we immediately get the full set of equations 
\begin{eqnarray}
\label{lhsDB}
\chi_{p+1} = (d + d^{\dagger}) A_{p}\,, ~~~~~~
(d+d^{\dagger}) \chi_{p+1}= \Box A_{p}
\end{eqnarray}
and inserting it into \eqref{lhsDA}, it yields
\begin{eqnarray}
\label{lhsDC}
\omega_{LHS} = \sum_{p=-\infty}^{\infty} Q \left(  c \gamma^{p-2} A_{p} \right) 
\end{eqnarray}
so the cohomology is trivial. To recover the SHS cohomology one has to additionally impose 
\begin{eqnarray}
\label{lhsDE}
\eta \omega =0
\end{eqnarray}
which selects the positive powers of $\gamma$ in \eqref{lhsDA} and therefore the cohomology computed as in previous subsections \ref{part-funct-2}, \ref{part-func-4}. 
Eventually, at the level of the partition function, counting the states yields (remember: $c \to q^{-2}c, \, \psi \to q \psi , \, \gamma \to q^{-1} \gamma)$ 
\begin{eqnarray}
\label{lhsDF}
P_{LHS}(q) &=& \frac{(1- q^{-2}) (1+q)^{n}}{1+ q^{-1}} +  \frac{(1- q^{-2}) (1+q)^{n}}{1+ q} - (1-q^{-2})(1+q)^{n} 
\nonumber \\
&=& (1-q^{-2})(1+q)^{n} \left( \frac{1}{1 + q^{-1}} + \frac{1}{1+q} -1\right) = 0 
\end{eqnarray}
where we used the identity $\sum_{p=-\infty}^{\infty} q^{p} =0$. The last term in the first line 
is needed to avoid double counting of the $\gamma$-independent state. As seen from the second eq.~in \eqref{lhsD}, in the Large Hilbert Space the picture 1 states are mapped by the operator $\theta(\beta)$ into picture 0 states, where each derivative of $\delta^{(p)}(\gamma)$ is associated to an inverse power of $\gamma$. 

Regarding the action $S[{\bf A}_0]$ \eqref{aaZC}:
\begin{eqnarray}
\label{lhsE} 
S[{\bf A}_{0}] = \frac12 (\omega, Q \omega)_{BV} &=&
\frac12 \int_{_{x, c, \gamma, \psi}} \omega_{cov} \wedge Q \omega_{cov} \, Y(c,\gamma) \nonumber \\
&=&  
\frac12 \int_{_{x, c, \gamma, \psi}}  \eta \left( \frac{c}{\gamma^{2}} \omega_{cov}\right) 
\wedge Q \omega_{cov} \,  
\end{eqnarray}
introducing 
${\widetilde \omega} = (1 + c \gamma^{-2}) \omega_{cov}$ allows us to rewrite the above action as follows 
\begin{eqnarray}
\label{lhsE} 
S[{\bf A}_{0}] = &=&  
\frac12 \int_{_{x, c, \gamma, \psi}}  \eta \left( \widetilde \omega\right) 
\wedge Q\widetilde \omega \,  
\end{eqnarray}
since $Q \widetilde \omega  = (1 + c \gamma^{-2})  Q \omega$. This is the 
starting point for the construction of Erler-Konopka-Sachs for superstring field theory \cite{Erler:2013xta}. 

\subsection{Interactions}

In this section we present a deformation of the free theory via allowed interacting term. Rather than wavefunctions of a Hilbert space, here we should focus on $\gamma, c$ as generators of an associative algebra, with values in the smooth functions of $x^\mu $ and $\psi^\mu$, which are naturally endowed with the associative multiplication of functions. A differential $Q = c \square + \gamma \psi^\mu \partial_\mu -\gamma^2 \partial_c$ is present. Therefore a BV bracket
\begin{equation}\label{BValgebra}
[\![\omega,\rho]\!] := (-1)^{\vert \omega\vert} \big(Q(\omega \rho) - Q(\omega) \rho - (-1)^{\vert \omega\vert} \omega Q(\rho)\big) \,, 
\end{equation}
where $\omega$ and $\rho$ are the functions $\omega=\omega(x,\psi,\gamma,c), \, \rho= \rho(x,\psi,\gamma,c)$, is at our disposal \cite{Gerstenhaber:1964}. For more on the topic of Gerstenhaber bracket and BV bracket, we invite to consult \cite{Kosmann-Schwarzbach:2000omw,Rogers2009,Fiorenza2011FormalityOK}. Since $Q^2=0$, then $Q$ is a derivation of this bracket. It can also be shown that the bracket satisfies the graded Jacobi identity.  Moreover the bracket is invariant under the BV pairing $(-,-)_{BV}$:
\begin{align*}
(\nu, [\![\omega,\rho]\!])_{{BV}} = & \,(-1)^{\vert \omega\vert}  \int [d^Dx d^D\psi dc d\gamma]  \, \nu \left(Q(\omega \rho) - Q(\omega) \rho - (-1)^{\vert \omega\vert} \omega Q(\rho)\right) Y \\
= & \, (-1)^{\vert \omega\vert}  \int [d^Dx d^D\psi dc d\gamma]  \, \left( (-1)^{\vert \nu\vert} (Q \nu) \omega \rho - (-1)^{2 \vert \omega\vert + \vert \nu\vert } Q(\nu \omega) - \nu Q\omega \rho \right) Y \\
= & \, (-1)^{\vert \omega\vert } \int [d^Dx d^D\psi dc d\gamma]  [\![\nu,\omega]\!] \rho \,  Y\\
= & (-1)^{\vert \omega\vert } ([\![\nu,\omega]\!], \rho)_{\text{BV}} \, .
\end{align*} 
Here we used that the measure and the picture changing operator $Y$ are BRST invariant.

With such a new bit added to the BRST operator, we can suggest an interaction term for the action functional and then check, by variational principle, what equations it satisfies and what symmetries the action has. 
So our object of interest is the following action:
\begin{equation}
\frac12 (\omega_{cov} , Q \omega_{cov} )_{{BV}} + \frac{1}{2} (\omega_{cov} , [\![\omega_{cov}, \omega_{cov}]\!])_{{BV}} \, .
\label{N1SFT}
\end{equation}
Before digging into its interaction term, a few comments are in order: since
\[
\omega_{cov} \equiv \sum_{p= 0}^{+\infty} \sum_{k=0}^{D} \gamma^p A_{p}(x)_{i_1, \dots i_k} \psi^{i_1} \dots \psi^{i_k} + c \sum_{p\geq 0}^{+\infty} \sum_{k=0}^{D} \gamma^p \chi_{p}(x)_{i_1, \dots i_k} \psi^{i_1} \dots \psi^{i_k} \, ,
\]
all of the terms in the second sums are in the kernel of $Y = c \delta'(\gamma)$ and the same is true after applying $Q$, therefore they drop out from the calculation. This affects also the linear gauge symmetry of the model.

A second useful consideration is that the degree of the covariant vertex (multiform) is therefore entirely carried by the $ \psi$'s. Using the notation $A_r$ to refer to the form degree of the object, we have therefore shown that:
\begin{align}
(\omega_{cov} , [\![\omega_{cov}, \omega_{cov}]\!])_{{BV}} =  (-1)^{p+r+1} \int \, & [d^D x\dots]   ((d+d^\dagger)A_r) \, A_pA_q + (-1)^r A_r \, ((d+d^\dagger)A_p) \, A_q \notag \\
& \; + (-1)^{r+p} A_r A_p ((d+d^\dagger)A_q) \; c \delta(\gamma) \, .
\label{interaction}
\end{align}
Therefore we obtain three terms with signs depending on the form degree. Singling out the top form ($D=4$) leads to
\begin{align}
&d^\dagger A_1 \, (2 A_4 A_0 + A_2 A_2)\,, ~~
(d A_0 + d^\dagger A_2) (2A_3 A_0 + 2 A_2 A_1), \nonumber \\
&(dA_1 + d^\dagger A_3) (2A_2 A_0), ~~(d A_2 + d^\dagger A_4) (2 A_1 A_0)\,, ~~ d A_3 (A_0A_0)  \, .
\end{align}
Some cancellations have occurred pairwise between the interaction terms in \eqref{interaction}, but the final bits in the Lagrangian are as above.


Recalling that
\begin{equation}
 \frac12 (\omega_{cov},Q \, \omega_{cov})_{\text{BV}}  =  \frac12 \int d^{D}x d^{D}\psi \, 
{\bf A}_{0} \left(d+d^{\dagger}\right) {\bf A}_{0}  \, ,
\end{equation}
we can now present the dynamical field equations. Let us start with the zero form field. We retrieve:
\begin{equation}\label{pivA}
d A_3 + A_4 d^\dagger A_1 + (d^\dagger A_4) A_1 + (d^\dagger A_2 )A_3 + A_2 d^\dagger A_3 = 0 .
\end{equation}
Then for the 1-form $A_1$ the field equation is:
\begin{equation}\label{pivB}
dA_2 + d^\dagger A_4 + d^\dagger (A_4 A_0 + \frac{1}{2} A_2A_2 ) - (d^\dagger A_2) A_2 - (d^\dagger A_4) A_0 = 0 \, .
\end{equation}
For completeness, we list the remaining field equations, for $A_2$, $A_3$ and $A_4$ respectively:
\begin{align}\label{pivC}
    & d A_1 + d^\dagger A_3 - d^\dagger(A_2A_1 + A_3A_0) + (d^\dagger A_3) A_0 + (d^\dagger A_1) A_2 + (d^\dagger A_2) A_1 = 0, \\\label{pivD}
    &d A_0 + d^\dagger A_2 + d^\dagger (A_2 A_0 ) - (d^\dagger A_2) A_0 = 0,\\\label{pivE}
    &d^\dagger A_1 - d^\dagger (A_1A_0) + (d^\dagger A_1) A_0 = 0.
\end{align}
In conclusion, for a generic form ${A}_{4-k}$:
\begin{equation}
(d {A}_{k-1} +d^\dagger {A}_{k+1}) - \frac{(-1)^k}{2} \sum_{l=0}^4 \left(d^\dagger ({A}_l {A}_{k-l+1}) - (d^\dagger {A}_l)  {A}_{k-l+1} - (-1)^l {A}_l d^\dagger  {A}_{k-l+1}\right) = 0 \, .
\end{equation}
Note that inside the sum we could substitute $d^\dagger$ with $d+d^\dagger$, since the differential is a derivation and therefore it will not contribute to the final result. Thus a BV bracket, with differential $d+d^\dagger$, is emerging. We will distinguish it from the previous BV bracket by a subscript and make use of it later in this section.

In order to study the consistency of the equations of motion at the non linear level, we first perform some easy checks and then we go ahead with the proof. 
Let us start with a consistency check on eqs. \eqref{pivA}-\eqref{pivE}. 
Let us re-write equation \eqref{pivE} as follows 
\begin{eqnarray}
    \label{cksD}
    &&* d *A_1 - * d *(A_1 A_0) + 
    (d^\dagger A_1) A_0 = 
     * d *A_1 - * d (*A_1 A_0) + 
    (d^\dagger A_1) A_0 \nonumber \\
    &&=
     * d *A_1 - * (d *A_1) A_0) - * (* A_1 d A_0)+  
    (d^\dagger A_1) A_0 \nonumber \\
    &&=* d *A_1 - d^\dagger A_1) A_0) - * (* A_1 d A_0)+  
    (d^\dagger A_1) A_0 = 
    * d *A_1 - * (* A_1 d A_0)  
   \end{eqnarray}
which implies 
\begin{eqnarray}
    \label{cksE}
    d *A_1 - (* A_1 d A_0) = 0  \,.  ~
    \end{eqnarray}
Now, we  can act with $d$ from the left and 
we get 
\begin{eqnarray}
    \label{cksF}
    - d (* A_1 \wedge d A_0) = - d(* A_1) \wedge d A_0 =  * A_1 \wedge d A_0 \wedge d A_0  =0
\end{eqnarray}
where we used \ref{cksE} again and $d A_0 \wedge d A_0 =0$\,. Therefore, the vanishing of right-side is consistent with the vanishing the left-hand side.

Let us consider now \eqref{pivD} and we write explicitly the third and fourth term as follows
\begin{eqnarray}
    \label{cksG}
    d A_0 + d^\dagger A_2 + * (* A_2 \wedge d A_0) = 0
\end{eqnarray}
and acting with $d^\dagger$ we get
\begin{eqnarray}
    \label{cksH}
    \square A_0 + * d (* A_2 \wedge d A_0) = 
    \square A_0 + * \Big( d (* A_2) \wedge d A_0\Big) = 0.
    \end{eqnarray}
Using the same eq. \eqref{cksG} in the form 
\begin{eqnarray}
    \label{cksI}
    d * A_2 = d A_0 \wedge * A_2 - * dA_0
\end{eqnarray}
we can recast \eqref{cksH} as follows 
\begin{eqnarray}
    \label{cksL}
     \square A_0 -  * \Big( d A_0 \wedge * d A_0\Big) = 
     \square A_0 -  \partial_\mu A_0 \partial^\mu A_0 = 0      
\end{eqnarray}
which implies the remarkable BV equation 
\begin{eqnarray}
    \label{cksM}
    e^{A_0} \square e^{- A_0} = 0 
\end{eqnarray}
in the perfect agreement with the BV algebra of the starting point \eqref{BValgebra}.  One can now perform a similar analysis on the other equations together with their gauge symmetries. 

For a complete check on the non-linear theory, we define the symbol $\mathcal{D} = d + d^\dagger$, such that $\mathcal{D}^2 = \square$. 
The derived bracket that we use is
\begin{equation}\label{BValgebraA}
[\![\omega,\rho]\!]_{\mathcal D} := (-1)^{\vert \omega\vert} \big({\mathcal D}(\omega \rho) - {\mathcal D}(\omega) \rho - (-1)^{\vert \omega\vert} \omega {\mathcal D}(\rho)\big) \,, 
\end{equation}
which satisfies the Gesternhaber algebra but it is not a BV algebra (because $\square \neq 0$) as shown in \cite{Kosmann-Schwarzbach:2000omw}.
Let us consider the equation of motion in the form
\begin{eqnarray}
    \label{DA1}
    {\mathcal D} {\bf A} + \frac12 [\![{\bf A},{\bf A}]\!]_{\mathcal D} =0
\end{eqnarray}
and acting with ${\mathcal D}$ on this equation from the left we get 
\begin{eqnarray}
    \label{DB1}
     {\mathcal D}^2 {\bf A} + \frac12  {\mathcal D} [\![{\bf A},{\bf A}]\!]_{\mathcal D} = 0 ,
\end{eqnarray}
where the main properties of the Gesternhaber algebra\footnote{We recall the basic properties that we use in the text: $Q$-compatibility \[
Q [\![\omega, \rho]\!] = [\![Q \omega , \rho]\!] + (-1)^{\vert \omega\vert -1} [\![\omega, Q\rho]\!],
\]
graded symmetry,
\[
[\![\omega, \rho]\!] = -(-1)^{(\vert \omega \vert -1)(\vert \rho\vert-1)} [\![ \rho, \omega]\!],
\]
and Jacobi identity
\[
[\![\omega , [\![\rho, \sigma]\!]]\!] = [\![[\![\omega, \rho]\!], \sigma]\!]+ (-1)^{(\vert \omega\vert -1) (\vert \rho \vert -1)} [\![\rho,[\![\omega, \sigma]\!]]\!].
\]
} are used. In addition, it has been used that the differential operator $\mathcal{D} = d + d^\dagger$ is of the second order because for all $A,B,C $ multiforms, it satisfies the identity 
\begin{eqnarray}
    \label{DBA}
    d^\dagger (A\wedge B\wedge C) = & d^\dagger (A\wedge B) \wedge C +(-1)^{\vert A \vert} A \wedge d^\dagger (B\wedge C) +(-1)^{\vert A\vert (\vert B \vert -1)} A \wedge d^\dagger (B \wedge C) \notag \\
    & - d^\dagger A \wedge  B \wedge C - (-1)^{\vert A\vert} A \wedge d^\dagger B \wedge C -(-1)^{\vert A \vert + \vert B \vert} A \wedge B \wedge d^\dagger C .
\end{eqnarray}
Using again the properties of the algebra, we finally put the result in the 
present form
\begin{eqnarray}
    \label{DC}
       \square {\bf A} + [\![{\mathcal D}{\bf A},{\bf A}]\!]_{\mathcal D} 
     = 
     \frac12 \left( {\mathcal D}^2 ({\bf A}\wedge {\bf A}) - 2 ({\mathcal D}^2 {\bf A}) \wedge {\bf A}\right)
\end{eqnarray}
but inserting \eqref{DA1} one arrives at:
     \begin{eqnarray}
    \label{DC}
    \square {\bf A} - \frac12 [\![ [\![{\bf A},{\bf A}]\!]_{\mathcal D},{\bf A}]\!]_{\mathcal D} 
     = 
     \frac12 \left( \square({\bf A}\wedge {\bf A}) - 2 (\square {\bf A}) \wedge {\bf A}\right).
\end{eqnarray}
The second term drops because of Jacobi identities for the Gerstenhaber bracket and the equation reads 
       \begin{eqnarray}
    \label{DD}
    \square {\bf A}
     = 
     \frac12 \left( \square({\bf A}\wedge {\bf A}) - 2 (\square {\bf A}) \wedge {\bf A}\right),
     \end{eqnarray}
which can be also put in the form 
\begin{eqnarray}
    \label{DF}
     \square {\bf A} -  \partial_\mu \mathbf{A} \wedge \partial^\mu \mathbf{A} = 
     e^{\bf A} \square e^{-\bf A} = 0
\end{eqnarray}
which is the consistency equation at the non-linear level. Notice that 
by setting the interaction to zero, we obtain the usual D'Alembertian equation. 
The interaction part measures the variation from being a derivation of 
$\mathcal D$. 

Now we want to address the question of the extra gauge symmetries of \eqref{N1SFT}. It turns out that, upon integration of the ghosts $c,\gamma$ which are carried by the PCO, it can be rewritten as:
\begin{equation}
  - \frac{1}{2}  \int_{x,\psi} \mathbf{A} (d+d^\dagger) \mathbf{A} + \sum_{k=0}^4 \frac{(-1)^{k}}{4} \int_{x,\psi} {A}_{4-k} \sum_{l=0}^4 (-1)^l [\![{A}_l, {A}_{k-l+1}]\!]_{\mathcal{D}} ,
  \label{N1SFTexpl}
\end{equation}
where indeed $[\![-, -]\!]_{\mathcal{D}}$ is \eqref{BValgebraA}, or else the bracket \eqref{BValgebra} with the differential operator $\mathcal{D} \equiv d + d^\dagger$ in place of $Q$.

{The algebraic structure of the interaction term then would suggest that the equation of motion
\eqref{DA1} has the symmetry
\begin{eqnarray}
\delta \mathbf{A} = (d+d^\dagger) {\bf \lambda} + [\![\mathbf{A}, {\bf \lambda}]\!]_{\mathcal{D}},
\label{gauge-extra}
\end{eqnarray}
as a consequence of the differential graded Lie algebra structure. 
{Before exploring this option, we would like to point out that the expression above 
is not an invariance of \eqref{pivC}. To understand this discrepancy we may recall that the invariance of the quadratic action \eqref{aaZC} becomes a BRST symmetry only due to the insertion of the PCO $Y$ in the measure, that is, 
\begin{eqnarray}
\label{GSA}
\delta \omega_{cov}Y= Q \Lambda Y 
\end{eqnarray}
However, since $Y$ is not a derivation of the bracket \eqref{BValgebraA}, this symmetry enhancement does not occur for the interaction term. } 

Let us now go back to the discussion of \eqref{gauge-extra}. As in the free case, the gauge parameters are not completely free (in the free case the gauge parameters ${\bf \lambda}$ have to be harmonic) and they have to satisfy some constraints. Computing the variation of the equations of motion 
we end up with the following complicate condition
\begin{eqnarray}
    \label{compA}
    {\mathcal D}^2 \lambda = 
    \frac12 \left( 
       {\mathcal D}^2 ({\bf A} \wedge {\bf \lambda}) - {\mathcal D}^2{\bf A} \wedge 
       {\bf \lambda} -  
      {\bf A} \wedge 
        {\mathcal D}^2 {\bf \lambda}
       \right) 
\end{eqnarray}
where the right hand side depends upon the multiform ${\bf A}$. If the latter is set to zero, we obtain the harmonicity condition on the multiform gauge parameters ${\mathbf \lambda}$. Note that the right-hand side expresses the fact that the differential operator $ {\mathcal D}^2$ is not a derivation of the wedge product. This equation can be drastically simplified to 
\begin{eqnarray}
    \label{compA}
    {\mathcal D}^2 \lambda = 
     \partial^\mu {\bf A} \wedge \partial_\mu {\bf \lambda}.
\end{eqnarray}
As an example, from 
eqn. \eqref{gauge-extra} we derive  the variation of the 1-form $A_1$,
\begin{align}
\delta A_1 = & \, d\lambda_0 + d^\dagger \lambda_2 \pm \big(d^\dagger (A_0 \lambda_2 + A_2 \lambda_0 + A_1 \lambda_1) - (d^\dagger A_0) \lambda_2 - A_0 d^\dagger \lambda_2 - (d^\dagger A_2) \lambda_0 - A_2 (d^\dagger \lambda_0)\notag \\
& - (d^\dagger A_1 ) \lambda_1 + A_1 d^\dagger \lambda_1\big).
\end{align}
A length, but not difficult computation shows that indeed component-by-components the equations of motion have this invariance with respect to constrained ghost fields. We deserve a deeper analysis in a future publication.






\subsection{Dimensional reduction}

In order to obtain the $N=1$ Hilbert space from the $N=2$ construction we recall that double the amount of supersymmetry implies the presence of to two sets of target space \emph{odd} bosons $\psi$, $\bar\psi$ as well as a doubling of their ghosts, with commutation relations:\[
\{\psi^\mu, \bar\psi^\nu\} = 2g^{\mu\nu} , \quad [\gamma, \bar\beta] = 1 = [\bar\gamma,\beta].
\] 
We use the more conventional representation of the Hilbert space with the two superghosts $\gamma$ and $\bar\gamma$ 
(in contrast with what has been used in \cite{Boffo:2024lwd}). 

Starting from the $N=2$ BRST charge, 
by a simple manipulation we can rewrite this expression as follows 
\begin{eqnarray}
    \label{dueA}
    Q_{N=2} &=& c P^2 + \gamma \bar\psi \cdot P + \bar\gamma\psi\cdot P  - \gamma\bar \gamma b\nonumber \\
     &=& c P^2 + \gamma (\bar\psi \cdot P + \psi\cdot P ) - \gamma^2 b
     + (\bar \gamma - \gamma) 
     (\psi\cdot P - \gamma b)  
     \nonumber \\
     &=& Q_{N=1} + \gamma_-
     Q_0 
\end{eqnarray}
where $Q_0 = (\psi\cdot P - \gamma b)$. It is easy to show that 
$Q_{N=1}^2=0$ (which is the nilpotency of the $N=1$ BRST charge). 
In addition, $Q_0^2=0$, which can be easily tested since there is no non-trivial commutation relation among those fields, and finally $\{Q_{N=1}, Q_0\}=0$. Therefore, the two charges form a double complex with two differental operators moving in the same Hilbert space. 

Now, we have two choices. One is the following. Compute the cohomology of $Q_0$ first and by the double complex theory compute the full cohomology. But that would lead to the full $Q_{N=2}$ cohomology already discussed in our paper. It would be interesting to compute the full cohomology as $H(Q_{N=1}, H(Q_0))$, but one has to take care of the overall ghost field $\gamma_-$. We deserve this analysis for future investigation. 

On the other side to achieve the dimensional reduction we can simply impose the condition 
\begin{eqnarray}
    \label{REDA}
    \gamma_- |0\rangle =0
\end{eqnarray}
namely we set to zero the $\gamma_-$ field, which means that we identify $\gamma$ with $\bar\gamma$ in the Hilbert space. That kills the second term of the BRST charge and finally the $N=2$ charge is reduced to the $N=1$. What about the fermions? 
It is sufficient to observe that 
on the Hilbert space, the combination $\psi_-$ is BRST invariant since it transforms into $\gamma_- P_\mu$. Therefore, we can freely set $\psi_- |0\rangle =0$ reducing the complete Hilbert space to the $N=1$ Hilbert space for which we have discussed the cohomology.  

\section{Spinorial Formulation}

\subsection{Relation with the Dirac Equation}
\label{correspondence}

An important question is: what is the relation with the previous results on the on-shell and off-shell formulation of the target space theory and the physical fermionic Dirac equation? To answer, we need to provide a dictionary between the two formulations. 

Given the multiforms 
\begin{eqnarray}
    \label{diA}
    {\bf A} = \sum_{p=0}^4 A_p = 
    \sum_{p=0}^4 \frac{1}{p!} A_{\mu_1 \dots \mu_p}  dx^{\mu_1} \wedge \dots \wedge dx^{\mu_p}  
\end{eqnarray}

Let us introduce some ingredients: 
we denote by $\Gamma^\mu$ the 
Dirac matrices with their Clifford algebra $\{\Gamma^\mu, \Gamma^\nu\} = 2 \eta^{\nu\nu}$; we denote $\Gamma_5$ the usual D=4 chirality operator, with $\{\Gamma_5, \Gamma_\mu\} =0$ and $\Gamma_5  = \frac{i}{4!} \epsilon_{\mu\nu\gamma\rho} \Gamma^\mu \Gamma^{\nu} \Gamma^{\gamma} \Gamma^{\rho}$; $\Gamma^{\nu\gamma\rho}$ is the totally antisymmetrized product of three Dirac matrices. 

Now, we can replace $1$-forms $dx^\mu$ with Gamma matrices as follows 
\begin{eqnarray}
    \label{diB}
    \Psi = \sum_{p=0}^4 \frac{1}{p!} A_{\mu_1 \dots \mu_p}  \Gamma^{\mu_1} \wedge \dots \wedge \Gamma^{\mu_p} = 
    \sum_{p=0}^4 \frac{1}{p!} A_{\mu_1 \dots \mu_p}  \Gamma^{\mu_1 \dots \mu_p} 
\end{eqnarray}
where $\Psi$ is bispinor. 
Now, the differential equations for ${\bf A}$ are translated in terms 
of $\Psi$ as a Dirac equation \cite{Ivanenko:1927ywm,Ivanenko:1985nq,Obukhov:1994sa} (also known as 
K\"ahler-Ivanenko-Dirac equation) 
\begin{eqnarray}
    \label{DA}
   (d + d^\dagger) {\bf A} =0 ~~~~~~ \longrightarrow  ~~~~~ i\Gamma^\mu \partial_\mu \Psi = 0\,. ~~~~~~~
\end{eqnarray}
The bispinor $\Psi$ transforms under the tensor product spinorial representation of the Lorentz symmetry. Note that the remaining index of the bispinor $\Psi$ in the Dirac equation is dummy. 

In terms of those fields we can define different currents, for example 
\begin{eqnarray}
    \label{DB}
    J_\mu = {\rm Tr} [\Psi \Gamma_\mu \Psi]\,, ~~~~~~~~~~~
    J^5_\mu ={\rm Tr} [\Psi \Gamma_5\Gamma_\mu \Psi]\,. ~~~~~~
\end{eqnarray}
where the trace is taken over the spinorial indices. 
If we compute the divergence of the currents, we get
\begin{eqnarray}
    \label{DC}
    \partial^\mu  J_\mu =  2  
    {\rm Tr}[\Psi \Gamma_\mu \partial^\mu\Psi] = 0 \, , ~~~~~~~~~~
    \partial^\mu  J^5_\mu =  
    {\rm Tr}[\Psi \Gamma_5\Gamma_\mu \partial^\mu\Psi] = 0 \, , 
\end{eqnarray}
because of the equations of motion \ref{DA}. We can reverse the implication: the conservation of the currents is equivalent to the Dirac equations of motion. 
These currents generate the $U_V(1)$ and $U_A(1)$ of the free fermionic free theory, eventually broken by anomalies at the quantum level. 

Now, we translate them into the form language 
\begin{eqnarray}
    \label{DD}
        J^{(1)} = {\rm Tr} [\Psi \Gamma_\mu \Psi] \, dx^\mu \,, ~~~~~~~
        J^{(3)} = {\rm Tr}[\Psi \Gamma_{\mu\nu\rho} \Psi]   \, 
        dx^\mu dx^\nu dx^\rho \, .
\end{eqnarray}
Using $\Gamma_{\mu\nu\rho} = i \epsilon_{\mu\nu\rho\sigma} \Gamma_5 \Gamma^\sigma$, we can re-express the chiral current $J^5_\mu$ in terms of the current $\bar \Psi \Gamma_{\mu\nu\rho} \Psi$. 
We recall that we can associate the $1$-form current 
$J_\mu dx^\mu$; this is not closed, but it is co-closed 
$d^\dagger J^{(1)} =0$.
The electric and the chiral charges for a $3$-dimensional surface are given by 
\begin{eqnarray}
    \label{DDA}
    Q(\Sigma_3)= \int_{\Sigma_3} J_3\,, ~~~~~~~~~
    Q_c(\Sigma_3)= \int_{\Sigma_1} \star J_1\,,
\end{eqnarray}
which are independent of the surface $\Sigma_3$.

\subsection{Anomalies}

Since in the previous sections we described a map from the multiform description to the Dirac theory, here we discuss one of the most important issues in the quantum theory of fermions, namely the presence of anomalies. As explained above, we have two currents $J_3$ and $J_1$ identified to the 
vector and axial-vector currents Tr$[\Psi \Gamma_\mu \Psi]$ and 
Tr$[\Psi \Gamma_5 \Gamma_\mu \Psi]$. At the quantum level computing the conservation laws of those currents (which are both conserved at the free level, since the Dirac theory is invariant under the corresponding rigid symmetries) one finds that they are broken by quantum corrections. In particular, for free theory of {\it chiral} fermions if one compute the correlation functions such as types 
\begin{eqnarray}
   {\mathcal A}_{\mu(\nu\rho)} =   \langle {\rm Tr}[\Psi \Gamma_5 \Gamma_\mu \Psi(x) ] {\rm Tr}[\Psi \Gamma_\nu \Psi(y)] {\rm Tr}[\Psi \Gamma_\rho \Psi(z)]\rangle 
\end{eqnarray}
one obtains that 
$\partial^\mu  {\mathcal A}_{\mu(\nu\rho)} \neq 0$ with $\partial^\nu {\mathcal A}_{\mu(\nu\rho)} =0$. This happens for any odd number of axial current insertion into a loop diagram. 
There are several ways to perform the computations, a very efficient one (Fujikawa method \cite{Fujikawa,Fujikawa2,Fujikawa3,Fujikawa4}) 
is to couple the two currents, the vector and the axial currents to background gauge fields as 
\begin{eqnarray}
    \label{anoA}
    S[B, B^5] = 
    \int d^4x \left( B^\mu{\rm Tr} [\Psi \Gamma_\mu \Psi] + 
    B^{5\mu} {\rm Tr}[\Psi \Gamma_\mu \Psi] \right) 
\end{eqnarray}
and then compute the partition function $\Gamma[B, B^5]$ integrating 
over the fermion fields (we use Euclidean signature) 
\begin{eqnarray}
    \label{anoAA}
e^{- \Gamma[B, B^5]} = \int {\mathcal D}\Psi 
e^{-S[\Psi] + S[B, B^5]}  
\end{eqnarray}
where ${\mathcal D}\Psi $ 
is the functional integration (properly regularized) over the Dirac fields and $S[\Psi]$ is the conventional Dirac action.
The classical action is invariant under the axial local symmetry 
\begin{eqnarray}
    \label{anoAB}
    \delta \Psi = i \eta(x) \Gamma_5 \Psi\,, ~~~~~~
    \delta B^5_\mu = \partial_\mu \eta(x) .
\end{eqnarray}
However, as is well-known, the measure ${\mathcal D}\Psi$ 
is not invariant under such a symmetry, in fact computing the Jacobian of the transformation 
\eqref{anoAB}, one gets the famous expression 
\begin{eqnarray}
   \label{anoAC}
   {\mathcal D}\Psi {\mathcal D}\bar \Psi ~~~~~\longrightarrow ~~~~
   {\mathcal D}\Psi {\mathcal D}\bar \Psi e^{ - 2i Tr \int d^4x \Gamma_5}
\end{eqnarray}
where the factor is ill-defined because of the vanishing of $Tr[\Gamma_5]$ and 
the divergence of $\int d^4x$ and needs a regularization. This can be performed by 
a Gaussian expression of the covariant Dirac operator $\slashed{\nabla}_B$ with respect to 
the gauge field $B_\mu$ coupled to the electric current. Without repeating the complete computation, one finds that Ward identities are modified into 
\begin{eqnarray}
    \label{anoAD}
    \partial_\mu \frac{\delta \Gamma[B, B^5]}{\delta B^5_\mu} = -\frac{1}{16 \pi^2} \epsilon_{\mu\nu\rho\sigma} \partial^\mu B^\nu \partial^\rho B^\sigma\,, ~~~~~~~~~~~
    \partial_\mu \frac{\delta \Gamma[B, B^5]}{\delta B_\mu} = 0\,. 
\end{eqnarray}
Of course, we can integrate the first equation by introducing a new term in the action \eqref{anoA} as follows 
\begin{eqnarray}
    \label{anoAE}
     S[B, B^5]  \longrightarrow  S[B, B^5] - \frac{1}{16\pi^2}
     \int d^4x \epsilon_{\mu\nu\rho\sigma}   B^{5 \mu} B^\nu \partial^\rho B^\sigma
\end{eqnarray}
but this modification will spoil the second Ward identity as 
\begin{eqnarray}
    \label{anoAF}
      \partial_\mu \frac{\delta \Gamma[B, B^5]}{\delta B_\mu} = \frac{1}{16\pi^2} 
        \epsilon_{\mu\nu\rho\sigma} \partial^\mu B^{5\nu} \partial^\rho B^\sigma.
\end{eqnarray}

Let us translate this in the language of multiforms. 
The action $S[B, B^5]$ becomes, with the substitution $B_\mu \rightarrow B_3, B^5_\mu \rightarrow B_1$,
\begin{eqnarray}
    \label{anoB}
    S[B_1, B_3] = \int (B_1 \wedge A_3 + B_3 \wedge A_1),
\end{eqnarray}
which, together with the Dirac action $S[\Psi]$, is invariant under the symmetry 
\begin{eqnarray}
    \label{anoC}
    \delta B_1 = d \lambda_0\,, ~~~~~~~~
    \delta B_3 = d^\dagger \lambda_4
\end{eqnarray}
where $\lambda_0, \lambda_4$ are gauge parameters (scalar and pseudoscalar). 

Therefore, we have the following Ward identities (again we assume to integrate over the quantum Dirac fields) 
\begin{eqnarray}
    \label{anoE}
    d \star   \frac{\delta \Gamma[B_1, B_3]}{\delta B_3(x)} = -\frac{1}{16 \pi^2} d B_1 \wedge d B_1\,, ~~~~~~
    d    \frac{\delta \Gamma[B_1, B_3]}{\delta B_1(x)} = 0\,, ~~~~~~
\end{eqnarray}
where we used $d \star$ instead of $d^\dagger$ for a better-looking equation. 

The main question is: in the language of multiforms, which symmetry gets an anomaly? 
To understand this, we observe the following: given the spinor $\Psi$ given in \eqref{diB}, satisfying the Dirac equations, we can multiply it by $\Gamma_5$ to get 
\begin{eqnarray}
    \label{bofA}
    \Psi \longrightarrow \Psi' = \Gamma_5 \Psi 
\end{eqnarray}
which still satisfy the Dirac equation. At the level of superform the multiplication with $\Gamma_5$ is equivalent to acting with $\ast$ Hodge dual on the components, namely we can redefine our field 
${\bf A}$ with ${\bf A}' = \ast {\bf A}$. This clearly reshuffles the various components 
without losing any physical information. This is a clear symmetry of the action since 
  $\ast (d + d^\dagger) \ast = d^\dagger + d$. 
Now, the anomaly emerges for the measure and we find from the computation of the Jacobian that (we refer to 
\cite{Alvarez-Gaume:1983ihn,Scrucca:1999uz} for a complete and exhaustive discussion) 
\begin{eqnarray}
   \label{anoG}
   {\mathcal D}{\bf A} ~~~~~\longrightarrow ~~~~
   {\mathcal D}{\bf A} e^{ - 2i Tr \int d^4x \ast}
\end{eqnarray}
where the factor is ill-defined and needs a regularization. Indeed, the 
trace of $\ast$ vanishes, this can be easily viewed by recalling that $\ast^2 = \pm 1$, which means that it separates the vector space in two eigenspaces with opposite eigenvalues, therefore the trace vanishes. Furthermore, the transformation is performed in the field space which is infinite dimensional, and this implies that the integral is divergent. This needs a regularization which can be performed by a Gaussian expression of the covariant operator $d_B + d^\dagger_B$ with respect to 
the gauge field $B_1$ coupled to the electric current (this is the reason that $B_1$ should couple to the electric current). Without repeating the complete computation, we find that Ward identities are anomalous as expected. Notice that equations \eqref{anoE} can be read as quantum Schwinger-Dyson functional equations more than Ward Identities associated to a symmetry, therefore the chiral anomaly can be viewed as an anomaly in the field redefinition more that an anomaly to a rigid symmetry. 

We delegate for future publication the analysis of the anomaly in presence of interaction such as studied above.

\section*{Acknowledgments}
 We would like to thank C.A. Cremonini, M. Serone, 
  and T. Voronov for very useful discussions. 
P.A.G. would like to thank LMU in Munich for an invitation where part of the work took place and is joined by E.B. in thanking the Mittag-Leffler Institute in Stockholm where the work has been completed during the program 
"Cohomological Aspects of Quantum Field Theory" 2025. E.B.~was supported by the GA\v{C}R grant PIF-OUT 24-10634O. I.S. is supported by the Excellence Cluster Origins of the DFG under Germany’s Excellence Strategy EXC-2094 390783311 and has benefited from hospitality at the Munich Institute for Astro-, Particle and BioPhysics (MIAPbP), also funded by the DFG under Germany´s Excellence Strategy – EXC-2094 – 390783311.

\bibliographystyle{unsrt}
\bibliography{N=1.bib}

\end{document}